\documentstyle[12pt,epsf,epsfig,a4]{article}
\begin{document}
\begin {titlepage}
\begin{flushleft}
FSUJ TPI QO-02/96
\end{flushleft}
\begin{flushright}
March, 1996
\end{flushright}
\vspace{20mm}
\begin{center}

{\Large \bf Homodyne detection for measuring coherent\\[1.ex]
 phase state distributions}
\vspace{15mm}
 
{\large \bf M. Dakna, L. Kn\"oll, and D.--G. Welsch}\\[1.ex]
{\large Friedrich-Schiller Universit\"at Jena}\\[0.5 ex]
{\large Theoretisch-Physikalisches Institut}\\[0.5 ex]
{\large Max-Wien Platz 1, D-07743 Jena, Germany}
\vspace{25mm}
\end{center}
\begin{center}
\bf{Abstract}
\end{center}
Using coherent phase states, parameterized phase state
distributions
for a single-mode radiation field are introduced and their integral
relation to the phase-parameterized field-strength distributions is
studied. The integral kernel is evaluated and the problem of direct
sampling of the coherent phase state distributions using balanced
homodyne detection is considered. Numerical simulations show that
when the value of the smoothing parameter is not too small the coherent
phase state distributions can be obtained with sufficiently well accuracy.
With decreasing value of the smoothing parameter the determination of
the coherent phase state distributions may be an effort, because both
the numerical calculation of the sampling function and the measurement
of the field-strength distributions are required to be performed with
drastically increasing accuracy.
\end{titlepage}
\renewcommand {\thepage} {\arabic{page}}
\setcounter {page} {2}

\section{Introduction}
\label{sec1}
The quantum-mechanical description of amplitude and phase quantities
and their measurement has turned out to be troublesome and is still of
matter of discussion (for a review, see Ref.~\cite{Lynch1}).
Attempts to introduce amplitude and phase operators
in quantum mechanics and particularly in the theory of the quantized
radiation field are nearly as old as quantum theory. Since Dirac's
introduction of amplitude and phase operators in 1927 \cite{Dirac1},
a series of concepts have been developed. Roughly speaking, there are
two routes to introduce amplitude and phase quantities and the associated
operators. In the first, amplitude and phase variables are introduced
by intrinsic phase operator definitions \cite{Dirac1,London1,Louisell1,%
Susskind1,Carruthers1,Garrison1,Turski1,Paul1,Levy1,Pegg1,Bergou1,%
Shapiro1,Vourdas1} or definitions closely related to phase space functions
\cite{Wodkiewicz1,Schleich1,Schleich2,Vogel1,Burak1,Garraway1,Schleich3}.
In the second, they are introduced by defining them more pragmatically
from the output observed in phase-sensitive measurements
\cite{Schubert1,Noh1}.

Phase measurements are usually based on some kinds of interference
experiments. To our knowledge the earliest measurements of phase
were carried out by Gerhardt, B\"{u}chler, and Litfin \cite{Gerhardt1}.
A powerful method pioneered by Walker and Caroll \cite{Walker1}
for phase measurements has been homodyne and heterodyne detection
techniques. These techniques, which were also used by Noh, Foug\`{e}res,
and Mandel \cite{Noh1} for defining measurement-assisted phase quantities,
have offered qualitatively new possibilities. In particular,
Smithey, Beck, Faridani, and Raymer \cite{Smithey1} have demonstrated
experimentally that balanced homodyne detection can be used to
reconstruct the quantum state of a signal-field mode (see also
Ref.~\cite{Smithey2}). The method is frequently called optical
homodyne tomography. The Wigner function of the signal mode
is reconstructed from the measured set of phase-parameterized
field-strength distributions \cite{Schubert2} using inverse Radon
transformation techniques, and the density matrix of the signal field is
then obtained by Fourier transforming the Wigner function.
It should be noted that tomographic methods have also been applied
to the reconstruction of quantum states of molecular vibrations
by Dunn, Walmsley, and Mukamel \cite{Dunn1}. Since the Wigner
function or the density matrix contains all knowable
information on the quantum statistics of the signal mode, they can
be used for inferring phase distributions that have been not directly
available from experiments, as was demonstrated by Beck, Smithey,
and Raymer \cite{Beck1} (see also Ref.~\cite{Smithey3}).

The tomographic method for reconstructing the Wigner function and the
density matrix of the signal mode from the phase-parameterized field-strength
distributions measured in balanced homodyne detection (for the detection
scheme see Refs.~\cite{Yuen1,Braunstein1,Vogel2}) is based on the fact
that knowledge of the field-strength distributions for all phases within
a $\pi$-interval is equivalent to knowledge of the quantum state
\cite{KVogel1}. In other words, the measured set of field-strength
distributions themselves can be regarded as being representative of the
quantum state of the signal mode and the question arises of whether
or not quantum-phase distributions can be sampled directly from
the measured field-strength distributions, without making the
detour via the density matrix. The problem is similar to those
related to the determination of the density matrix itself.
Whereas in the first experiments the density matrix was obtained
via the reconstructed Wigner function, recently methods of direct sampling
of the density matrix from the measured data have been
suggested and successfully applied \cite{Kuehn1,Vogel3,Ariano1,Munroe1,%
Ariano2,Leonhardt1,Ariano3,Richter1,Leonhardt2,Zucchetti1}.
The feasibility of measuring quantities by direct sampling
depends on the relationships between the quantities and the field-strength
distributions recorded. In particular, these relationships should be stable
against sampling errors, so that measurement of a sufficiently large
set of data should enables one to obtain the quantities with sufficiently
well precision. Clearly, different concepts of introduction of
phase variables lead to different phase distributions and to different
relations between them and the field-strength distributions.
In the present paper we restrict attention to the class of quantum-phase
distributions that can be defined on the basis of the coherent phase states
\cite{Levy1,Shapiro1,Vourdas1}. Roughly speaking, these
distributions are smoothed London phase state distributions
which can be be parameterized with respect to the degree of smoothing.
The connection of the coherent phase state (CPS) distributions with
the phase-parameterized field-strength distributions is considered and
the problem of direct sampling of the CPS distributions from the
field-strength distributions is studied. Numerical simulations are carried
out to illustrate the applicability of the method.

In Sec.~\ref{sec2} the CPS distributions are introduced. Their integral
relation to the phase-parameterized field-strength distributions is
studied in Sec.~\ref{sec3}. Results of a computer simulation of
direct sampling of CPS distributions are reported in Sec.~\ref{sec4},
and a summary and some concluding remarks are given in Sec.\ref{sec5}.
 
\section{Coherent phase state distributions}
\label{sec2}
In the long history of the quantum phase problem
the eigenstates of the ``exponential-phase'' operator
\begin{equation}
\label{2.1}
       \hat{E}_{-} = \sum_{n=0}^\infty | n \rangle \langle n+1 |
\end{equation}
($|n\rangle$, photon-number states) have been played a central role.
In particular, the states
\begin{equation}
\label{2.2}
    \big| e^{i\phi} \big\rangle = \sum_{n=0}^\infty e^{in\phi} |n\rangle
\end{equation}
satisfying the eigenvalue equation
\begin{equation}
\label{2.3}
     \hat{E}_{-} \, \big| e^{i\phi} \big\rangle
     = e^{i\phi} \, \big| e^{i\phi} \big\rangle
\end{equation}
have been referred absolutely to as phase states. These states
introduced by London \cite{London1} resolve the unity, but they
are non-orthogonal and non-normalizable. As was shown by
L\`{e}vy-Leblond \cite{Levy1}, the eigenvalue equation
\begin{equation}
       \hat{E}_{-} \, | z \rangle = z \, | z \rangle
\label{2.4}
\end{equation}
has also other solutions than those in Eq.~(\ref{2.2}), viz.
\begin{equation}
\label{2.5}
|z\rangle = \sqrt{1-|z|^2} \,
\sum_{n=0}^{\infty} z^{n} |n\rangle, \quad |z| < 1.
\end{equation}
These normalizable states, which form an over-complete basis,
have been re-examined and called coherent phase states
by Shapiro and Shepard \cite{Shapiro1}. Their properties
have also been studied by Vourdas \cite{Vourdas1} who showed
that they are SU(1,1) Perelomov coherent states $| z;k \rangle$
corresponding to the $k$ $\!=$ $\!\textstyle{\frac{1}{2}}$ representation
\cite{Perelomov1}. In the limit $|z|$ $\!\to$ $\!1$ the states
$|z\rangle$ become the non-normalizable London states in Eq.~(\ref{2.2}):
\begin{equation}
\label{2.6}
\big|e^{i\phi}\big\rangle
= \lim_{\epsilon\to 0}\frac{1}{\sqrt{1-e^{-2\epsilon}}} |\phi,\epsilon \rangle,
\end{equation}
where the notations
\begin{equation}
\label{2.7}
|\phi,\epsilon \rangle \equiv |z\rangle, \quad
z = e^{-\epsilon} e^{i\phi} \quad (\epsilon>0)
\end{equation}
have been introduced.

The states $|\phi,\epsilon\rangle$ can be used to define
$\epsilon$-parameterized phase state distributions -- CPS distributions --
of a radiation-field mode via the overlap of its quantum state
with coherent phase states:
\begin{equation}
\label{2.8}
p(\phi,\epsilon)
= N^{-1}(\epsilon) \,
\langle \phi,\epsilon |\,\hat{\varrho}\,| \phi,\epsilon\rangle,
\end{equation}
where
\begin{equation}
\label{2.9}
N(\epsilon)
= \int_{-\pi}^{\pi} {d}\phi \,
\langle \phi,\epsilon |\,\hat{\varrho}\,| \phi,\epsilon\rangle
\end{equation}
($\hat{\varrho}$, density matrix of a radiation-field mode).
In the limit $\epsilon$ $\!\to$ $\!0$ the distributions
$p(\phi,\epsilon)$ become the familiar London phase state
distribution $p(\phi,0)$ $\!=$ $\langle e^{i\phi}|\hat{\varrho}
|e^{i\phi}\rangle$. The CPS distributions $p(\phi,\epsilon)$
can be regarded as smoothed London phase state distributions,
which can be directly sampled from the field-strength distributions
measured in balanced homodyne detection, as we will see below.

\begin{figure}
\centering\epsfig{figure=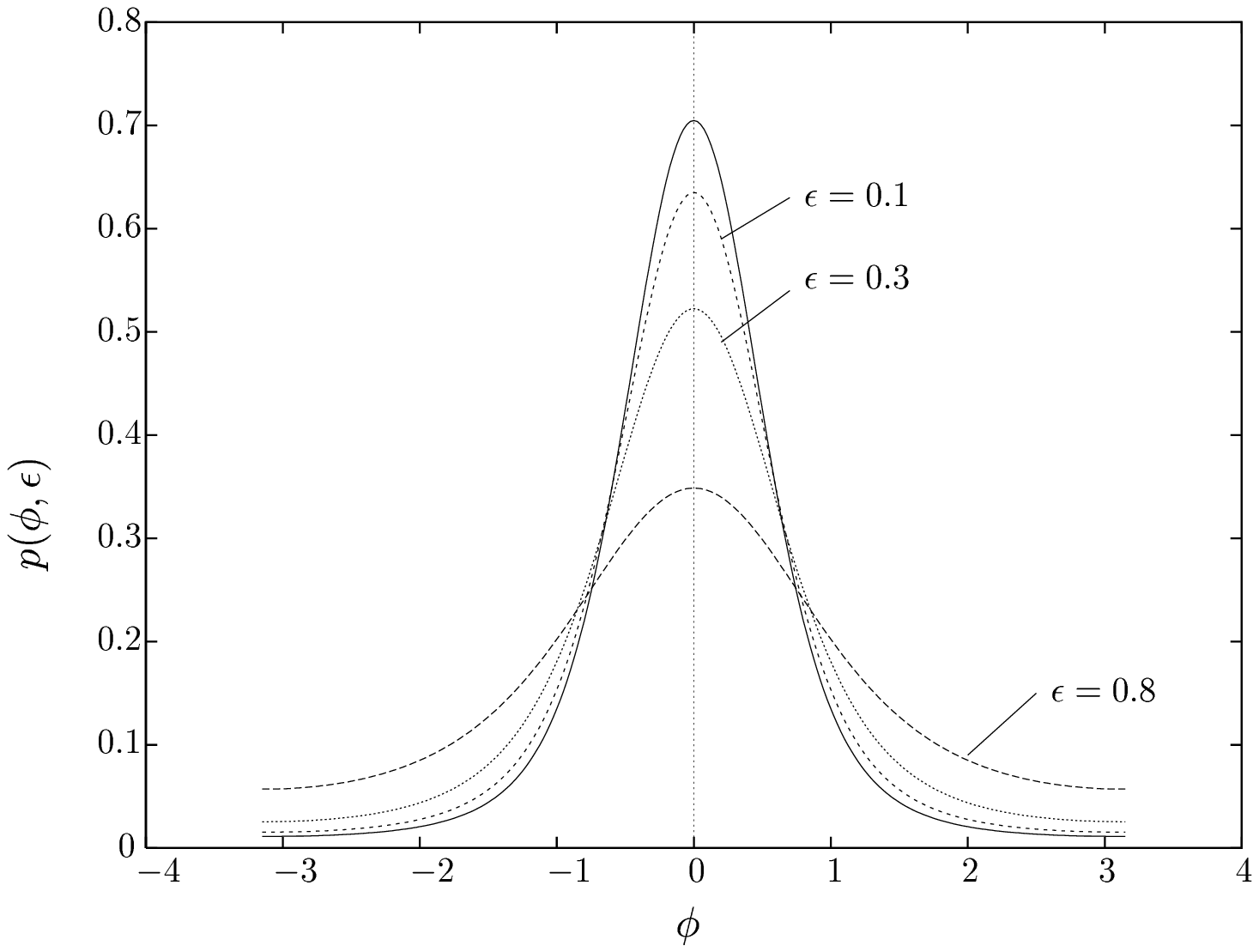,width=\linewidth}
\caption{ \label{fig:figr1}
The CPS distributions $p(\phi,\epsilon)$
for a coherent state with mean photon number
$\langle\hat{n}\rangle=1$ are shown for various values
of the smoothing parameter $\epsilon$. In the limit when
$\epsilon$ goes to zero $p(\phi,\epsilon)$ approaches
the London phase state distribution (solid line).}
\end{figure}

\begin{figure}
\centering\epsfig{figure=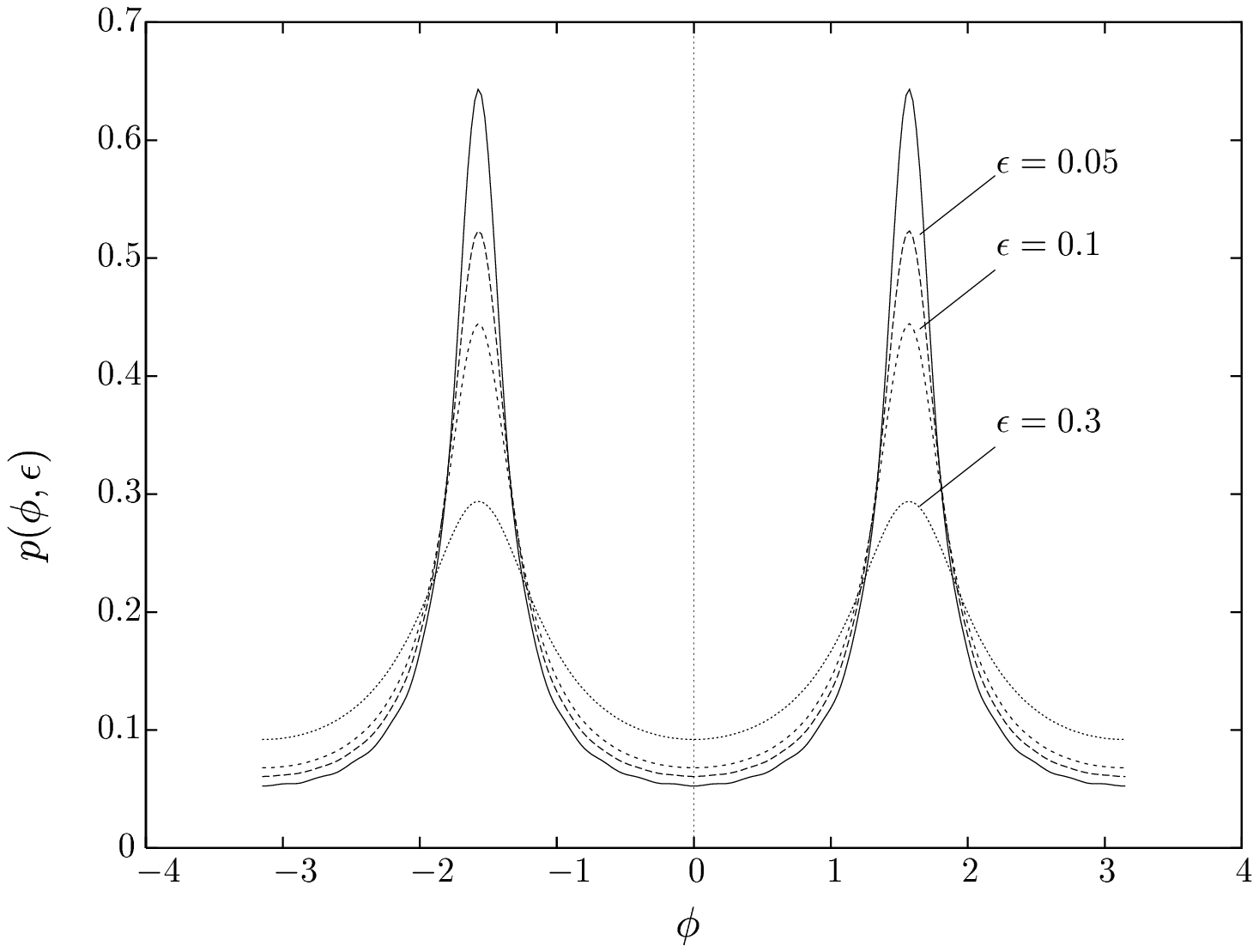,width=\linewidth}
\caption{ \label{fig:figr2}
The CPS distributions $p(\phi,\epsilon)$
for a squeezed vacuum state with mean photon number
$\langle\hat{n}\rangle=1$ are shown for various values
of the smoothing parameter $\epsilon$. In the limit when
$\epsilon$ goes to zero $p(\phi,\epsilon)$ approaches
the London phase state distribution (solid line).}
\end{figure}

\begin{figure}
\centering\epsfig{figure=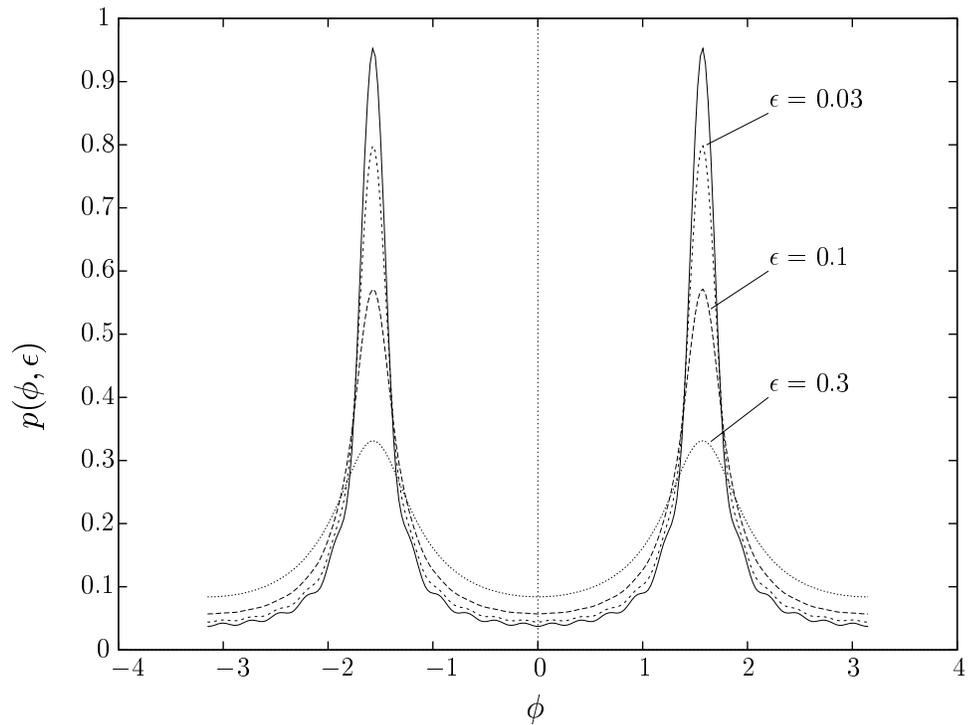,width=\linewidth}
\caption{ \label{fig:figr3}
The CPS distributions $p(\phi,\epsilon)$
for a squeezed vacuum state with mean photon number
$\langle\hat{n}\rangle=2$ are shown for various values
of the smoothing parameter $\epsilon$. Note that compared
with Fig.2 smaller values of $\epsilon$
are needed in order to obtain CPS distributions that are
comparably close to the London phase distribution (solid line).}
\end{figure}

To illustrate the effect of the smoothing parameter $\epsilon$,
in Figs.~\ref{fig:figr1} -- \ref{fig:figr3} CPS distributions and
the London phase state distributions of coherent and squeezed
states are plotted for various values of the coherent amplitude
and the squeezing parameter. When the value of the smoothing
parameter is decreased the CPS distributions become closer and closer
to the corresponding London phase state distribution, which is observed,
at least in tendency, in the limit $\epsilon$ $\!\to$ $\!0$. From
comparison of Figs.~\ref{fig:figr2} and \ref{fig:figr3} we further see that
with increasing (mean) photon number the value of the smoothing parameter
is required to be decreased in order to observe CPS distributions that
are comparably close to the London phase state distribution.
This can be explained as follows.
In the expansion in photon-number states of the London phase states
the weight of the states with large photon numbers is increased
with increasing (mean) number of photons in the state under study. In
the CPS distributions the overlap of these states with the state under
study would be suppressed when the smoothing parameter would be unchanged.

\section{Relation to the field-strength distributions}
\label{sec3}
The phase-parameterized field-strength distributions measurable
in perfect balanced homodyning,
\begin{equation}
p({\cal F},\varphi)=
\langle {\cal F},\varphi |\,\hat{\varrho}\,| {\cal F},\varphi\rangle,
\label{3.01}
\end{equation}
were introduced by Schubert and Vogel \cite{Schubert2}
(for details see, e.g., \cite{Vogel3}). In Eq.~(\ref{3.01}), the
$|{\cal F},\varphi\rangle$ are the eigenvectors of a field-strength
operator $\hat{F}(\varphi)=F \hat{a}+F^\ast\hat{a}^\dagger$
at chosen phase $\varphi$ ($F$ $\!=$ $\!|F| e^{-i\varphi}$,
$\hat{a}^\dagger$ and $\hat{a}$ are the photon creation and destruction
operators, respectively).
It should be noted that non-perfect detection introduces additional
noise, so that the measured distributions are, in general,
smeared field-strength distributions. These are convolutions of the
true field-strength distributions with Gaussians \cite{Vogel2}:
\begin{equation}
p({\cal F},\varphi;s) = \int {d}{\cal F} \, p({\cal F}',\varphi)\,
p({\cal F}\!-\!{\cal F}';s), 
\label{3.02}
\end{equation}
where
\begin{equation}
p({\cal F};s) = \left(2\pi |s| |F|^2\right)^{-\frac{1}{2}}
\exp\!\left(-\frac{{\cal F}^2}{2 |s| |F|^2}\right)
\label{3.03}
\end{equation}
and $s$ $\!=$ $\!1\!-\!\eta^{-1}$, $\eta$ being the detection
efficiency ($\eta$ $\!\leq$ $\!1$).

The relation between the field-strength distributions and the
CPS distributions may be obtained as follows.
Using the expansion of the density operator $\hat{\varrho}$
in $s$-parameterized displacement operators \cite{Cahill1}
and relating the averages of the $s$-parameterized
displacement operators to the characteristic functions of the
field-strength distributions \cite{KVogel1} yields
(see, e.g., Refs.~\cite{Leonhardt1,Ariano3})
\begin{equation}
\hat{\varrho} =  \int_0^\pi {d}\varphi \int {d} {\cal F}
\,p({\cal F},\varphi;s) \, \hat{K}({\cal F},\varphi;-s),
\label{3.3}
\end{equation}
where the operator integral kernel $\hat{K}({\cal F},\varphi;-s)$
reads as
\begin{equation}
\hat{K}({\cal F},\varphi;-s)
=\frac{|F|^2}{\pi} \int {d} y \, |y|\,
\exp\!\left\{iy\left[\hat{F}(\varphi)\!-\!{\cal F}\right]
-{\textstyle\frac{1}{2}}sy^2|F|^2\right\}\!.
\label{3.4}
\end{equation}
Note that Eq.~(\ref{3.4}) can formally be applied also to cases where the
detection efficiency is less than unity. Combining Eqs.~(\ref{2.8})
and (\ref{3.3}), we find that $p(\phi,\epsilon)$ can be related
to $p({\cal F},\varphi)$ as
\begin{equation}
p(\phi,\epsilon) = N^{-1}(\epsilon)
\int_{0}^{\pi}d\varphi\int_{-\infty}^{\infty} {d}{\cal F}\,
p({\cal F},\varphi;s)\, K_\epsilon(\phi,{\cal F},\varphi;s),
\label{3.8}
\end{equation}
where
\begin{equation}
K_\epsilon(\phi,{\cal F},\varphi;s)
= \big\langle\phi,\epsilon\big|\hat{K}({\cal F},\varphi;-s)
\big|\phi,\epsilon\big\rangle.
\label{3.8a}
\end{equation}

\begin{figure}
\noindent
\begin{minipage}[b]{0.5\linewidth}
\centering\epsfig{figure=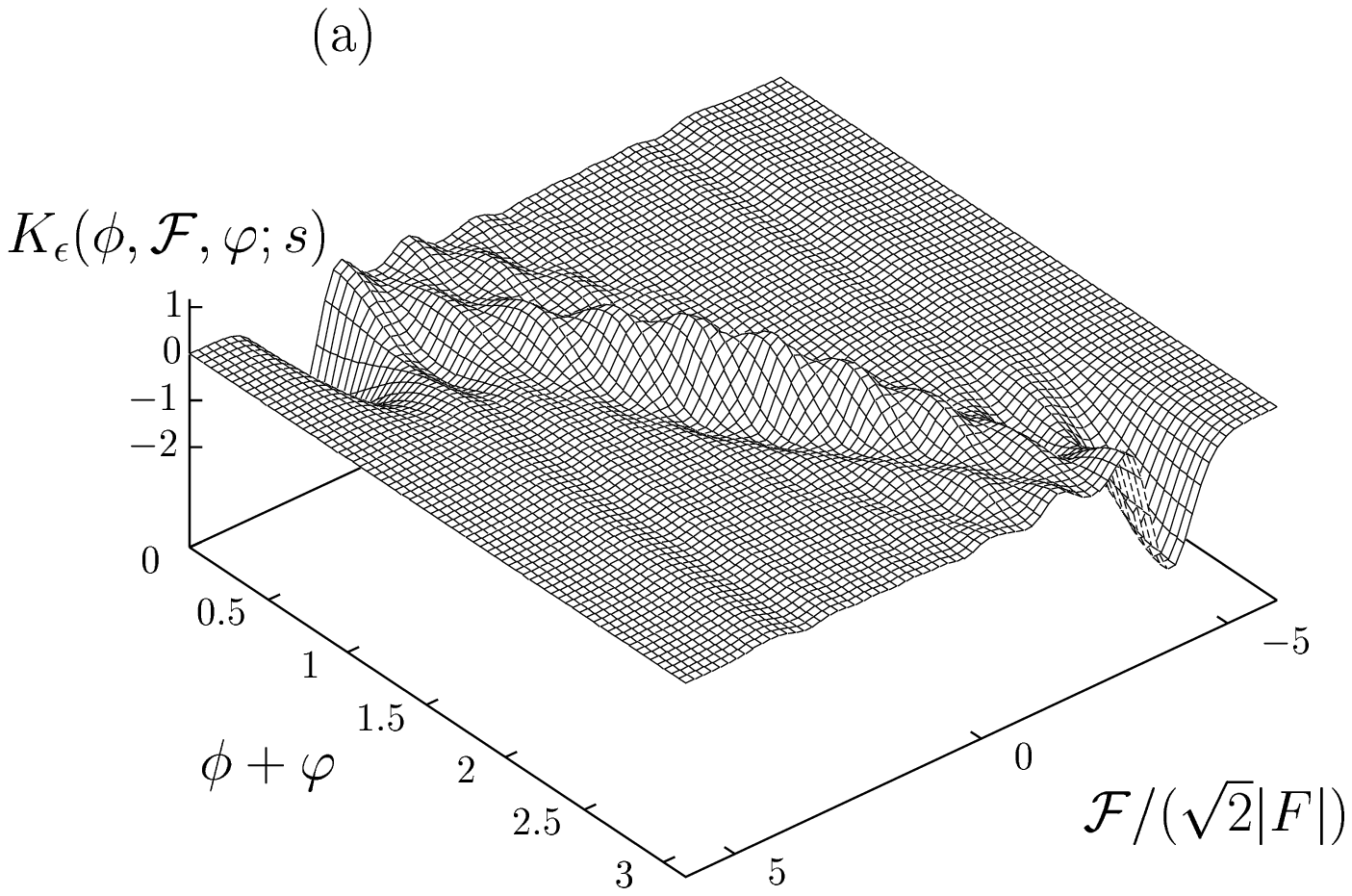,width=\linewidth}
\end{minipage}\hfill
\begin{minipage}[b]{0.5\linewidth}
\centering\epsfig{figure=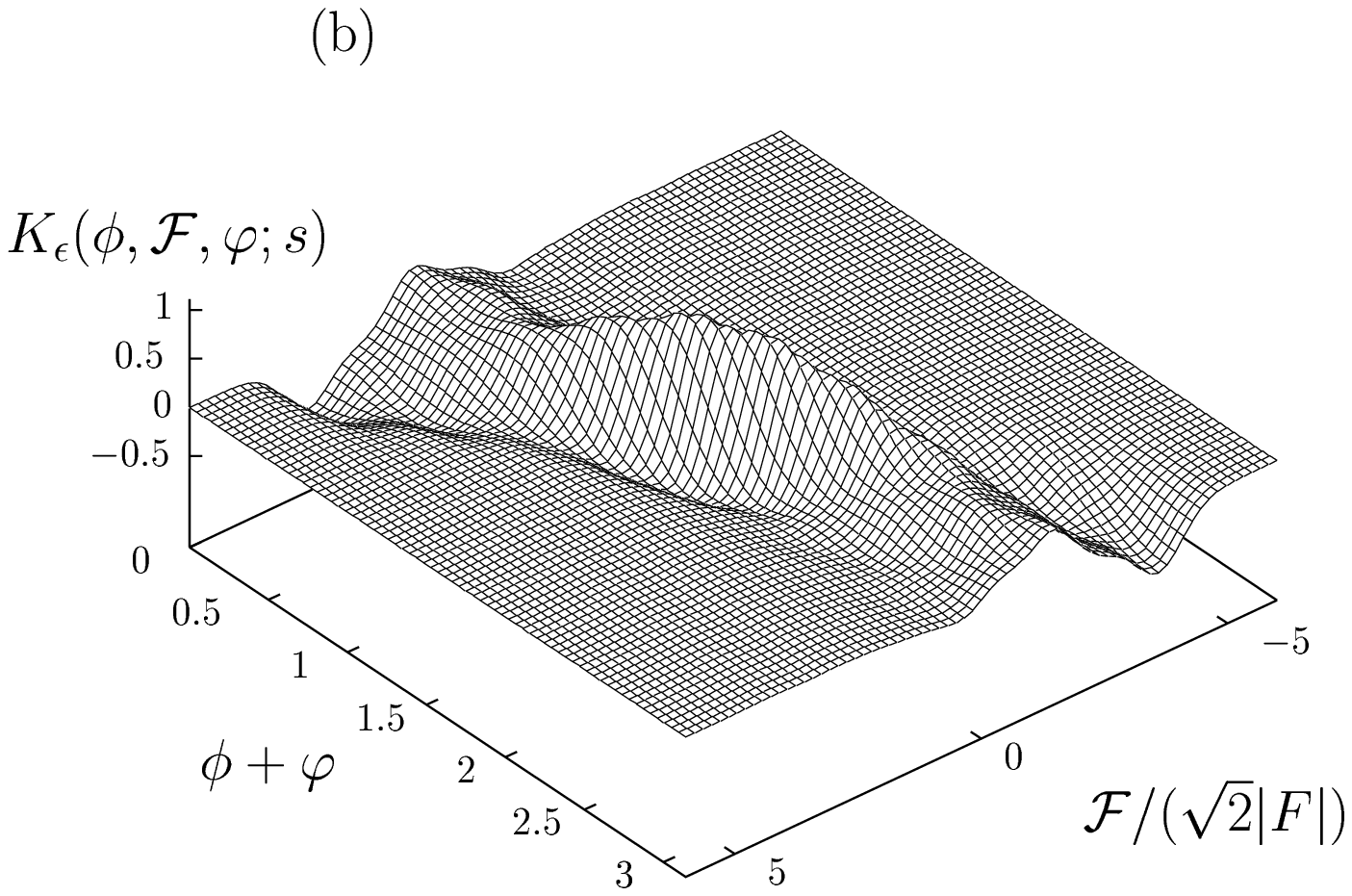,width=\linewidth}
\end{minipage}

~

\vspace{3cm}

~

\begin{minipage}[b]{0.5\linewidth}
\centering\epsfig{figure=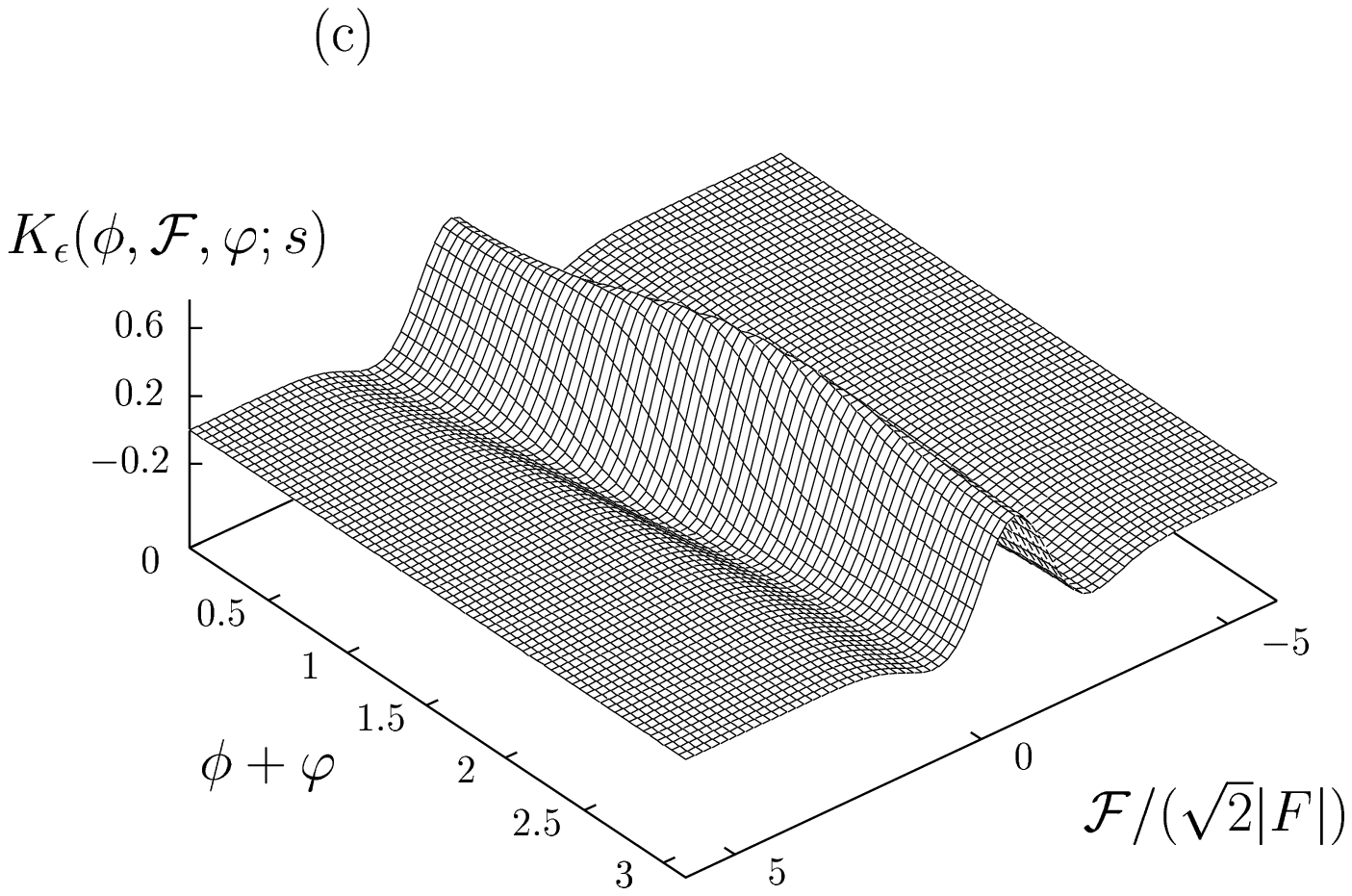,width=\linewidth}
\end{minipage}\hfill
\begin{minipage}[b]{0.5\linewidth}
\centering\epsfig{figure=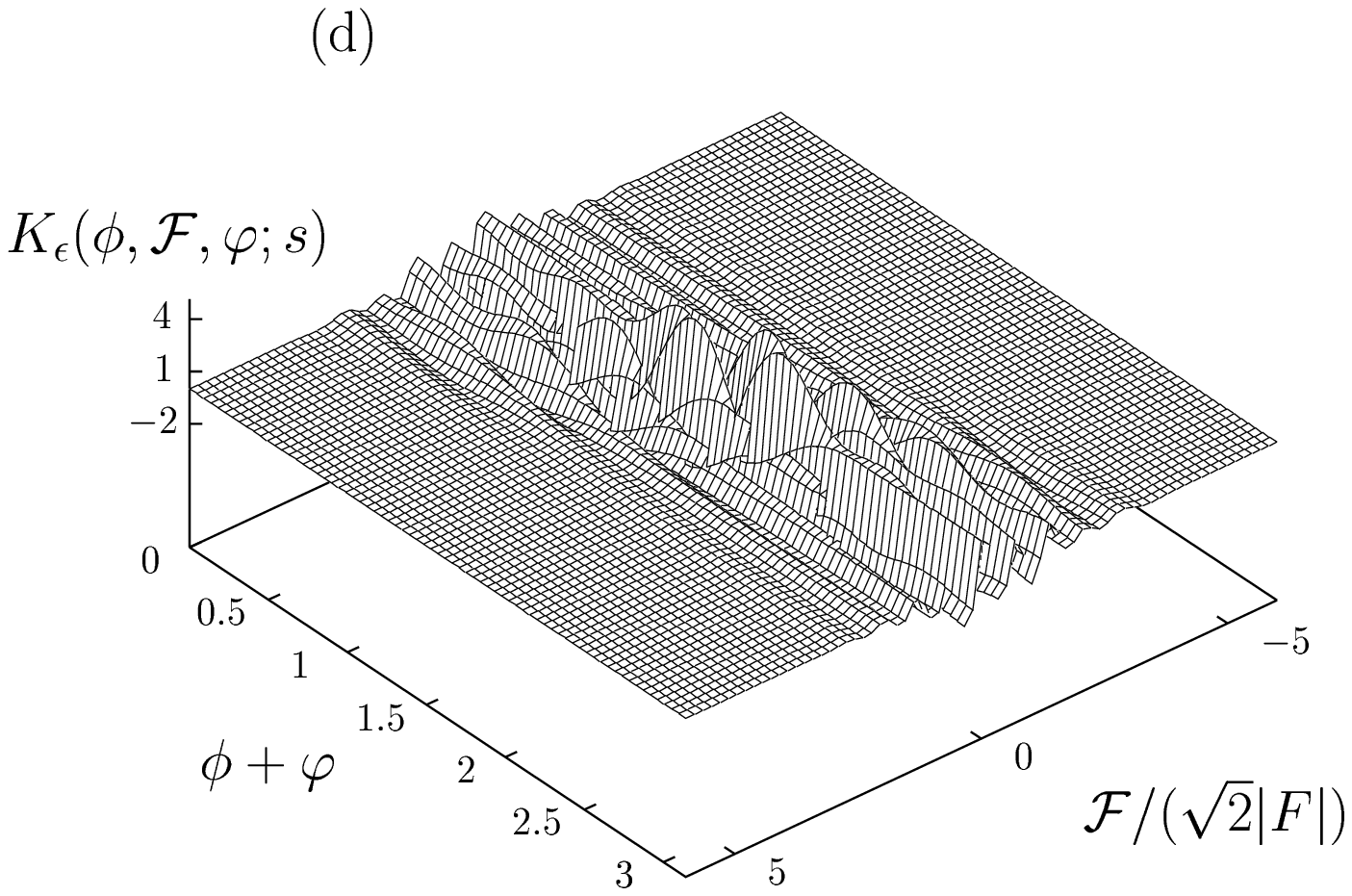,width=\linewidth}
\end{minipage}
\caption{ \label{fig:figr4}
$K_\epsilon(\phi,{\cal F},\varphi,;s)$ is shown
as a function of ${\cal F}$ and $\phi\!+\!\varphi$ for
various values of the smoothing parameter $\epsilon$ and the
detection efficiency $\eta$. \protect\\
(a) $\epsilon$ $\!=$ $\!0.1$, $\eta$ $\!=$ $\!1$ ($s$ $\!=$ $\!0$).
\protect\\
(b) $\epsilon$ $\!=$ $\!0.3$, $\eta$ $\!=$ $\!1$ ($s$ $\!=$ $\!0$).
\protect\\
(c) $\epsilon$ $\!=$ $\!0.8$, $\eta$ $\!=$ $\!1$ ($s$ $\!=$ $\!0$).
\protect\\
(d) $\epsilon$ $\!=$ $\!0.1$, $\eta$ $\!=$ $\!0.8$ ($s$ $\!=$ $\!-0.25$).}
\end{figure}

Equation~(\ref{3.8}) can be regarded as the basis equation for
direct sampling of CPS distributions from the difference-count statistics
in balanced homodyning. Let us consider the sampling function
$K_\epsilon(\phi,{\cal F},\varphi;s)$ in more detail.
Expanding the $|\phi,\epsilon\rangle$ in photon-number states [see
Eq.~(\ref{2.5})], Eq.~(\ref{3.8a}) can be rewritten as
\begin{equation}
K_\epsilon(\phi,{\cal F},\varphi;s)
 = (1-e^{-2\epsilon})
\sum_{n=0}^{\infty}\sum_{m=0}^{\infty}
K_{nm}({\cal F},\varphi;-s)
e^{i(n-m)\phi}
e^{-\epsilon(n+m)},
\label{3.9}
\end{equation}
where
\begin{equation}
K_{nm}({\cal F},\varphi;-s) =
\langle n|\hat{K}({\cal F},\varphi;-s)|m\rangle
\label{3.10}
\end{equation}
is the sampling function for determining the density matrix in
the photon-number basis, which has been studied in detail in a
number of papers \cite{Ariano1,Munroe1,Ariano2,Leonhardt1,Ariano3,%
Richter1,Leonhardt2}. It is given by
\begin{equation}
K_{nm}({\cal F},\varphi;-s) = e^{i(n-m)\varphi}f_{nm}(x;s),
\label{3.10a}
\end{equation}
$x={\cal F}/(\sqrt{2}|F|)$, where $f_{nm}(x;s)$ may be expressed
in terms of parabolic cylinder functions. From Eqs.~(\ref{3.9}) and
(\ref{3.10a}) we see that $K_\epsilon(\phi,{\cal F},\varphi;s)$ is
a $2\pi$-periodic function of the sum phase $\phi$ $\!+$ $\!\varphi$.
Since the relation $f_{nm}(x;s)$ $\!=$ $f_{mn}(x;s)$ is valid,
$K_\epsilon(\phi,{\cal F},\varphi;s)$ is an even function of the
sum phase. For $\eta$ $\!>$ $\!1/2$ the function $f_{nm}(x,s)$
is bonded. For large values of $x$ it behaves
like $x^{-{|n-m|-2}}$, so that $K_\epsilon(\phi,{\cal F},\varphi;s)$
is bounded for  $\epsilon$ $\!>$ $\!0$. In Figs.~\ref{fig:figr4}(a) --
\ref{fig:figr4}(d) $K_\epsilon(\phi,{\cal F},\varphi;s)$ is shown for various
values of the smoothing parameter $\epsilon$ and the detection
efficiency $\eta$.

Let us first restrict attention to perfect detection ($\eta$ $\!=$ $\!1$,
i.e., $s$ $\!=$ $\!0$), Figs.~\ref{fig:figr4}(a) -- \ref{fig:figr4}(c).
We see that with increasing $\epsilon$ the fine structure in the
dependence on ${\cal F}$ of $K_\epsilon(\phi,{\cal F},\varphi;s)$ is lost.
The fine structure obviously results from highly oscillating terms in
Eq.~(\ref{3.9}), i.e., from terms with $n,m$ $\!\gg$ $\!1$. These terms
become more and more suppressed when $\epsilon$ is increased. On the
other hand, the width of the interval of ${\cal F}$ in which
$K_\epsilon(\phi,{\cal F},\varphi;s)$ is essentially nonzero
slowly decreases with increasing $\epsilon$. Apart from finenesses,
the CPS distributions may therefore be expected to reflect typical
properties of the London phase state distribution even for values
of $\epsilon$ that are not extremely small. Clearly, for $\epsilon$
$\!\to$ $\!0$ the calculation of $K_\epsilon(\phi,{\cal F},\varphi;s)$
becomes an effort, because the number of terms in Eq.~(\ref{3.9})
drastically increases. From Fig.\ref{fig:figr4}(d) we see that in
the case of non-perfect detection ($\eta$ $\!<$ $\!1$, i.e.,
$s$ $\!<$ $\!0$) the function $K_\epsilon(\phi,{\cal F},\varphi;s)$
becomes highly structurized, so that increasing effort has to be made
to sample CPS distributions from the field-strength distributions.

In general, the CPS distributions $p(\phi,\epsilon)$ are determined
by the field-strength distributions $p({\cal F},\varphi)$ in more or
less extended intervals of ${\cal F}$ and $\varphi$. In particular,
Figs.~\ref{fig:figr4}(a) -- (c) clearly show that the intervals of ${\cal F}$
and $\varphi$ $\!+$ $\!\phi$ in which $K_\epsilon(\phi,{\cal F},\varphi;s)$
is essentially nonzero are not concentrated in the vicinity
of the points ${\cal F}$ $\!=$ $\!0$ and $\varphi$ $\!+$ $\!\phi$
$\!=$ $\!\pi/2$, respectively. Hence, the CPS distributions
(and the London phase state distribution) cannot be obtained from
the field-strength distributions at ${\cal F}$ $\!=$ $\!0$ and
$\varphi$ $\!+$ $\!\phi$ $\!=$ $\!\pi/2$. This contradicts the
operational phase distribution $p(\phi)$ $\!\propto$ $\!p({\cal F}\!=\!0,
\varphi\!=\!\pi/2\!-\!\phi)$ introduced by Vogel and Schleich
\cite{Vogel1}.

\section{Direct sampling of coherent phase state distributions}
\label{sec4}
\begin{figure}
\centering\epsfig{figure=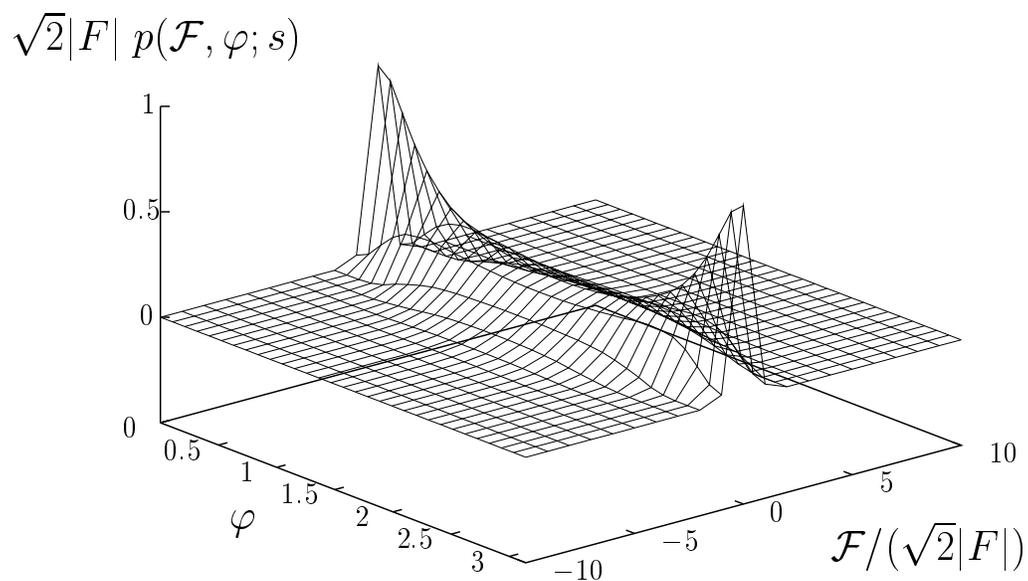,width=\linewidth}
\caption{\label{fig:figr5}
The measured set of field-strength distributions
$p({\cal F},\varphi;s)$ for a squeezed vacuum state with mean photon
number $\langle\hat{n}\rangle$ $\!=$ $\!1$ is shown. It is obtained
from a computer simulation, where perfect detection ($\eta$ $\!=$ $\!1$,
i.e., $s$ $\!=$ $\!0$) is considered and 30$\times$10$^4$ events 
at 30 phase values are assumed to be recorded
(the measured, discrete points are linked by straight lines).}
\end{figure}

\begin{figure}
\centering\epsfig{figure=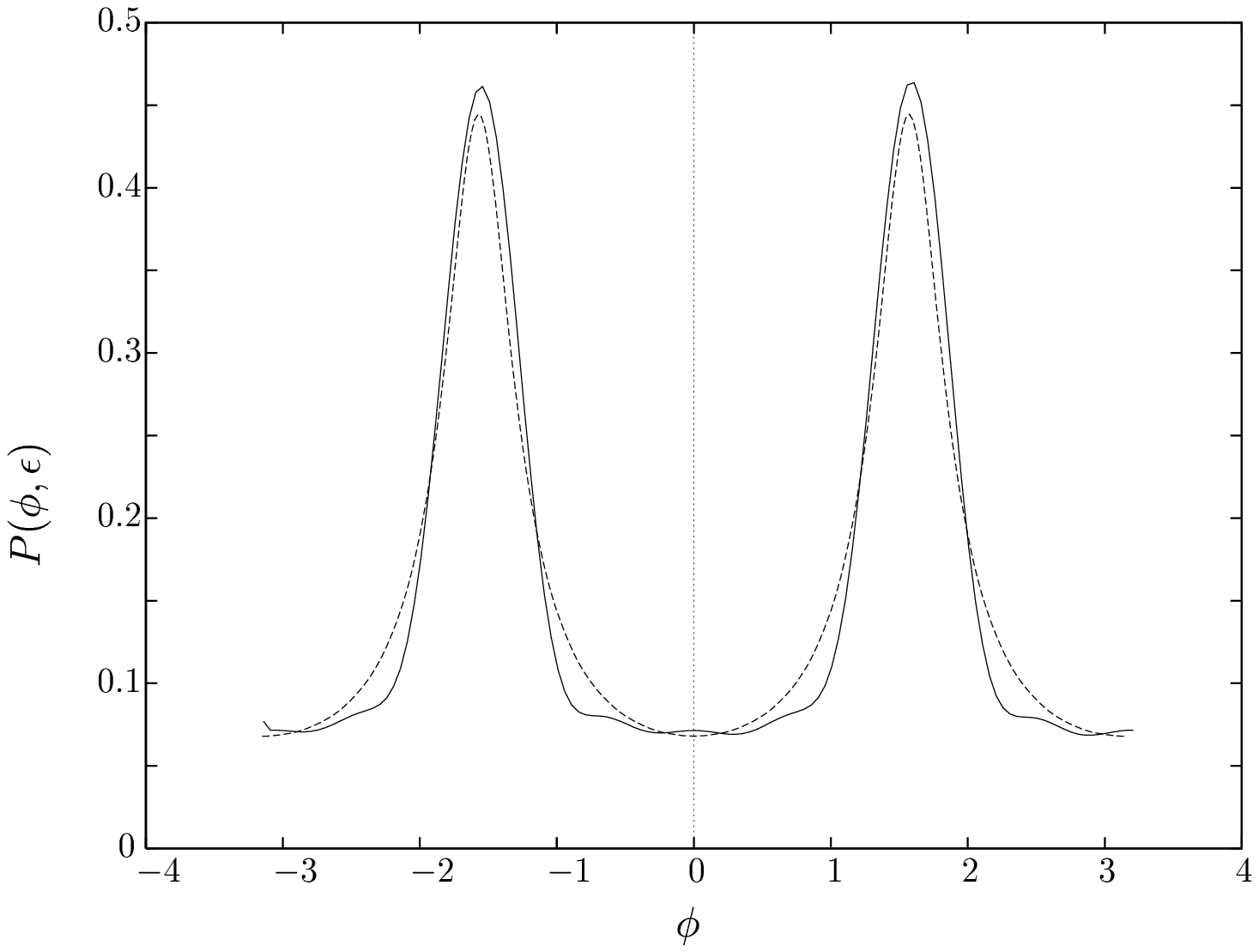,width=\linewidth}
\caption{\label{fig:figr6}
The CPS distribution for $\epsilon$ $\!=$ $\!0.1$
which is directly sampled from the measured field-strength 
distributions plotted in Fig.5 is shown
and compared with the theoretical curve from
Fig.2 (dashed line).}
\end{figure}

As already mentioned, Eq.~(\ref{3.8}) can be used for direct
sampling of CPS distributions from the field-strength distributions
measured in balanced homodyne detection. To give an example,
let us consider the
determination of CPS distributions for a squeezed vacuum state.
The result of a computer simulation of measurement of a
set of field-strength distributions for a squeezed vacuum state
with mean photon number $\langle\hat{n}\rangle$ $\!=$ $\!1$ is
shown in Fig.~\ref{fig:figr5}. In the simulation perfect detection
is considered and 30$\times$10$^4$ events at 30 phase values
are assumed to be recorded.
The feasibilty of direct sampling of CPS distributions $p(\phi,\epsilon)$
from the measured field-strength distributions is demonstrated in
Fig.~\ref{fig:figr6} for $\epsilon=0.1$.

Comparing the sampled CPS distribution with the calculated
distribution, we see that the measured double-peak structure
is in good agreement with the theoretical prediction. In particular,
in the intervals of phase in which reduced phase fluctuations
are observed a sufficiently well agreement between theory and experiment
is found. Relative large discrepancies between the measured and the
calculated curves are seen in the intervals of phase in which enhanced
phase fluctuations are observed. Clearly a better agreement and a higher
accuracy can be achieved when the number of phases at which the
field-strength distributions are measured and the number of events
recorded at each phase are increased.

\section{Summary and concluding remarks}
\label{sec5}
We have studied the integral relation of the CPS distributions
$p(\phi,\epsilon)$ to the phase-parameterized field-strength
$p({\cal F},\varphi)$ measurable in balanced homodyne
detection and calculated the corresponding integral kernel.
Since the CPS distributions
can be regarded as smoothed London phase state distributions,
the degree of smoothing being determined by the
parameter $\epsilon$ in the coherent phase states
$|z\!=\!e^{-\epsilon}e^{i\phi}\rangle$ ($\epsilon$ $\!>$ $\!0$),
in the limit $\epsilon$ $\!\to$ $\!0$ the London phase state
distribution is observed.

With decreasing value of the smoothing parameter $\epsilon$
the information about the phase statistics is increased, because
the smoothing parameter $\epsilon$ controls the interval of
the field strength used for ``probing'' the phase statistics
of the state under consideration. With decreasing value of
$\epsilon$ this interval is increased and comprises the
whole field-strength axis in the limit $\epsilon$ $\!\to$ $\!0$.
Hence, with decreasing value of $\epsilon$ the determination
of the integral kernel requires inclusion in the calculation
of increasing values of the field strength. This
point must be considered very carefully when the state under
consideration contains large field-strength values that are
desired to be included in an analysis of the phase statistics.

In the paper we have calculated the integral kernel
using an expansion in the photon-number basis. Since in this
basis with decreasing value of $\epsilon$ the values of the photon
numbers which must be taken into account increase, highly oscillating
functions are involved in the expansion and much effort must
be made to obtain results with reasonable accuracy.
This difficulty might be overcome by using consequently
a field-strength basis and avoiding the detour via the
photon-number basis, as has recently been demonstrated for
direct sampling of the density matrix in a field-strength basis
\cite{Zucchetti1}.

The integral relation between the CPS distributions and the
field-strength distributions can be used for direct sampling
of the CPS distributions from the data measured in balanced
homodyning, the (bounded) integral kernel playing the role
of the sampling function. To illustrate the method, we have
performed a computer simulation of measurement of a set
of field-strength distributions and sampling from them the
CPS distributions for a squeezed vacuum state with
mean photon number $\langle\hat{n}\rangle$ $\!=$ $\!1$.
In this case a smoothing parameter of about 0.1 is sufficient
to detect the features typical for the phase distribution
of the state. Using realistic number of phases at which the
field-strength distribution is measured and realistic numbers
of events recorded the sampled CPS distributions are found to 
be in good agreement with the calculated ones.

\section{Acknowledgments}
We are grateful to T. Opatrn\'{y} for stimulating discussions.
This work was supported by the Deutsche Forschungsgemeinschaft.

\newpage
\newpage
\newpage
\newpage
\newpage

\begin{thebibliography}{99}

\bibitem{Lynch1}
R. Lynch,
Phys. Rep. {\bf 256}, 367 (1995).
\bibitem{Dirac1}
P.A.M. Dirac,
Proc. R. Soc. Lond. A{\bf 114}, 234 (1927).
\bibitem{London1}
F. London,
Z. Phys. {\bf 37}, 915 (1926),
ibid. {\bf 40}, 193 (1927).
\bibitem{Louisell1}
W.H. Louisell,
Phys. Lett. {\bf 7}, 60 (1963).
\bibitem{Susskind1}
L. Susskind and J. Glogower,
Physics {\bf 1}, 49 (1964).
\bibitem{Carruthers1}
P. Carruthers and M.M. Nieto,
Phys. Rev. Lett. {\bf 14}, 387 (1965),
Rev. Mod. Phys. {\bf 40}, 411 (1968).
\bibitem{Garrison1}
J.C. Garrison and J. Wong,
J. Math. Phys. {\bf 11}, 2242 (1970).
\bibitem{Turski1}
L.A. Turski,
Physica {\bf 57}, 432 (1972).
\bibitem{Paul1}
H. Paul,
Fortschr. Phys. {\bf 22}, 657 (1974).
\bibitem{Levy1}
J.-M. L\`{e}vy-Leblond,
Ann. Phys. {\bf 101}, 319 (1976).
\bibitem{Pegg1}
D.T Pegg and S.M. Barnett,
Europhys. Lett. {\bf 6}, 483 (1988),
Phys. Rev. A{\bf 39}, 1665 (1989).
\bibitem{Bergou1}
J. Bergou and B.G. Englert,
Ann. Phys. {\bf 209}, 479 (1991).
\bibitem{Shapiro1}
J.H. Shapiro and S.R. Shepard,
Phys. Rev. A{\bf 43}, 3795 (1991).
\bibitem{Vourdas1}
A. Vourdas,
Physica Scripta T{\bf 48}, 84 (1993).
\bibitem{Wodkiewicz1}
K. W\'{o}dkiewicz,
Phys. Rev. Lett. {\bf 52}, 1064 (1984),
Phys. Lett. {\bf 115}A, 304 (1986).
\bibitem{Schleich1}
W.P. Schleich, D.F. Walls, and J.A. Wheeler,
Phys. Rev. A{\bf 38}, 1177 (1988).
\bibitem{Schleich2}
W. Schleich, R.J. Horowicz, and S. Varro,
Phys. Rev. A{\bf 40}, 7405 (1989).
\bibitem{Vogel1}
W. Vogel and W. Schleich,
Phys. Rev. A{\bf 44}, 7642 (1991).
\bibitem{Burak1}
D. Burak and K. W\'{o}dkiewicz,
Phys. Rev. A{\bf 46}, 2744 (1992).
\bibitem{Garraway1}
B.M. Garraway and P.L. Knight,
Phys. Rev. A{\bf 46}, R5346 (1992),
Physica Scripta T{\bf 48}, 66 (1993).
\bibitem{Schleich3}
W. Schleich, A. Bandilla, and H. Paul,
Phys. Rev. A{\bf 45}, 6652 (1992).
\bibitem{Schubert1}
M. Schubert,
Phys. Lett. {\bf 27}A, 698 (1968).
\bibitem{Noh1}
J.W. Noh, A. Foug\`{e}res, and L. Mandel,
Phys. Rev. Lett. {\bf 67}, 1426 (1991),
ibid. {\bf 71}, 2579 (1993),
Phys. Rev. A{\bf 45}, 424 (1992), ibid. A{\bf 46}, 2840 (1992),
A{\bf 47}, 4535, 4541 (1993).
\bibitem{Gerhardt1}
H. Gerhardt, U. B\"{u}chler, and G. Litfin,
Phys. Lett. {\bf 49}A, 119 (1974).
\bibitem{Walker1}
N.G. Walker and J.E. Caroll,
Electron. Lett. {\bf 20}, 981 (1984),
Opt. Quant. Electron. {\bf 18}, 355.
\bibitem{Smithey1}
D.T. Smithey, M. Beck, A. Faridani, and M.G. Raymer,
Phys. Rev. Lett. {\bf 70}, 1244 (1993).
\bibitem{Schubert2}
M. Schubert and W. Vogel,
Phys. Lett. {\bf 68}A, 321 (1978).
\bibitem{Smithey2}
D.T. Smithey, M. Beck, J. Cooper, M.G. Raymer, and A. Faridani,
Physica Scripta T{\bf 48}(1993) 35.
\bibitem{Dunn1}
T.J. Dunn, I.A. Walmsley, and S. Mukamel,
Phys. Rev. Lett. {\bf 74}, 884 (1995).
\bibitem{Beck1}
M. Beck, D.T. Smithey, and M.G. Raymer,
Phys. Rev. A{\bf 48}, R890 (1993).
\bibitem{Smithey3}
D.T. Smithey, M. Beck, J. Cooper, and M.G. Raymer,
Phys. Rev. A{\bf 48}, 3159 (1993).
\bibitem{Yuen1}
H.P. Yuen and J.H. Shapiro,
IEEE Trans. Inform. Theor. {\bf 26}, 78 (1980).
\bibitem{Braunstein1}
S. Braunstein,
Phys. Rev. A{\bf 42}, 474 (1990).
{\bf 5} (1993) 65
\bibitem{Vogel2}
W. Vogel and J. Grabow,
Phys. Rev. A{\bf 47}, 4227 (1993).
\bibitem{KVogel1}
K. Vogel and H. Risken,
Phys. Rev. A{\bf 40}, 2847 (1989).
\bibitem{Kuehn1}
H. K\"{u}hn, D.-G. Welsch, and W. Vogel,
J. Mod. Opt. {\bf 41}, 1607 (1994).
\bibitem{Vogel3}
W. Vogel and D.-G. Welsch,
{\em Lectures on Quantum Optics} (Akademie-Verlag, Berlin, 1994).
\bibitem{Ariano1}
G.M. D'Ariano, C. Macchiavello, and M.G.A. Paris,
Phys. Rev. A{\bf 50}, 4298 (1994),
Phys. Lett. {\bf 195}A, 31 (1994).
\bibitem{Munroe1}
M. Munroe, D. Boggavarapu, M.E. Anderson, and M.G. Raymer,
Phys. Rev. A{\bf 52}, R924 (1995).
\bibitem{Ariano2}
G.M. D'Ariano, U. Leonhardt, and H. Paul,
Phys. Rev. A{\bf 52}, R1801 (1995),
ibid. A{\bf 52}, 4899 (1995).
\bibitem{Leonhardt1}
U. Leonhardt, H. Paul, and G.M. D'Ariano,
Phys. Rev. A{\bf 52},4899 (1995).
\bibitem{Ariano3}
G.M. D'Ariano,
Quantum Semiclass. Opt. {\bf 7}, 693, (1995).
\bibitem{Richter1}
Th. Richter, Phys. Lett. A, in press.
\bibitem{Leonhardt2}
U. Leonhardt, M. Munroe, T. Kiss, M.G. Raymer, and Th. Richter,
Opt. Commun., in press.
\bibitem{Zucchetti1}
A. Zucchetti, W. Vogel, M. Tasche, and D.-G. Welsch,
Prep\-rint FSUJ TPI Q-01/96 (Fried\-rich-Schil\-ler-Uni\-ver\-si\-t\"at Jena,
Theo\-re\-tisch-Phy\-sikalisches Institut, Jena, 1996).
\bibitem{Perelomov1}
A.M. Perelomov,
Commun. Math. Phys. {\bf 26}, 222 (1972),
Funct. Anal. Appl. {\bf 7}, 225 (1973).
\bibitem{Cahill1}
K.E. Cahill and R. Glauber,
Phys. Rev. {\bf 177}, 1857 (1969).
\end{thebibliography}
\end{document}